\documentclass[prl,aps,twocolumn]{revtex4-1}
\usepackage{graphicx} 
\usepackage[usenames]{color}
\usepackage{amsmath,amssymb}
\usepackage{gensymb}
\usepackage{natbib}
\usepackage{color}
\def\strutdepth{\dp\strutbox}
\def\nw#1{\strut\vadjust{\kern-\strutdepth\vtop to0pt{\vss\hbox to\hsize
{\hskip\hsize\hskip5pt$\leftarrow$\hss\strut}}}{\em #1}}

\begin{document}

\title{Capillary Pressure and Contact Line Force on a Soft Solid }
\author{Antonin Marchand$^1$, Siddhartha Das$^2$, Jacco H. Snoeijer$^2$ and Bruno Andreotti$^1$.}
\affiliation{
$^{1}$Physique et M\'ecanique des Milieux H\'et\'erog\`enes, UMR
7636 ESPCI -CNRS, Univ. Paris-Diderot, 10 rue Vauquelin, 75005, Paris\\
$^{2}$Physics of Fluids Group and Mesa+ Institute, University of Twente, P.O. Box 217, 7500 AE Enschede, The Netherlands
}

\date{\today}%

\begin{abstract}
The surface free energy, or surface tension, of a liquid interface gives rise to a pressure jump when the interface is curved. Here we show that a similar capillary pressure arises at the interface of soft solids. We present experimental evidence that immersion of a thin elastomeric wire into a liquid induces a substantial elastic compression due to the solid capillary pressure at the bottom. We quantitatively determine the effective surface tension from the elastic displacement field, and find a value comparable to the liquid-vapor surface tension. Most importantly, these results also reveal the way the liquid pulls on the solid close to the contact line: the capillary force is not oriented along the liquid-air interface, nor perpendicularly to the solid surface, as previously hypothesized, but towards the interior of the liquid.%
\end{abstract}

\maketitle

Surfaces of crystalline solids can be shaped by surface stresses \cite{MS04}. These stresses induce phenomena as surface reconstruction \cite{BGIE97,FF97}, surface segregation \cite{WK77}, surface adsorption \cite{I04}, elastic instabilities \cite{MS04}, self assembly \cite{AVMJ88,MSM05}, and nanostructuration \cite{R02}. On the contrary, much less is known about surface stress or surface tension in soft amorphous materials, as gels and elastomers. A simple physical picture is that these materials are essentially liquid-like, with a small elastic modulus to resist shear deformations~\cite{MZK02}. Can such soft solids be shaped by capillary forces, just like ordinary liquids? Recent experiments provide evidence that this is indeed possible~\cite{SG87,CGS96,MPFPP10,JXWD11}. A thin filament of solid gel was observed to exhibit a Rayleigh-Plateau instability \cite{MPFPP10}: analogous to liquid jets,  surface variations appear in order to lower the surface free energy.
Similarly, Jerison et al. \cite{JXWD11} demonstrated that deformations of an elastic film by a liquid drop can only be explained quantitatively by accounting for the free energy of the solid surface. They argued that one should include an additional stress due to the curvature of the solid that is induced by the presence of the liquid.
This raises the intriguing prospect of a \emph{solid capillary pressure}, arising when a solid-liquid interface is curved: can it be measured, what is its magnitude, and what are its physical consequences? 

\begin{figure}[t!]
\includegraphics{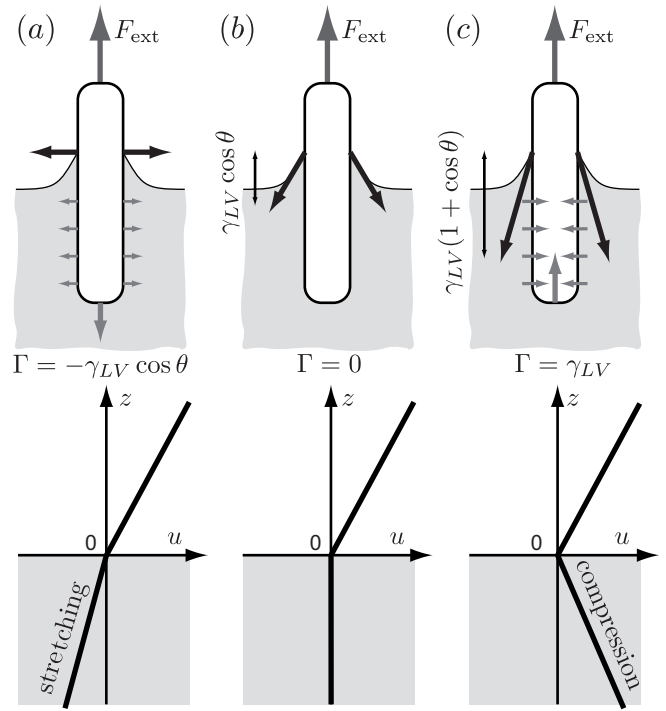}
\vspace{-6mm}
\caption{Top: Scenarios for distribution of capillary forces on an elastic wire partially immersed in a liquid~\cite{GBQ04,L61,R78,SG87,CGS96,LAL96,PBBB08,PRDBRB07,NRGB07,DMAS11}. The liquid-on-solid force near the contact line (black arrows) could be oriented (a) normal to the wire, (b) along the liquid-vapour interface, (c) pointing into the liquid phase. Recovering the total thermodynamic force (per unit contact line), $F_{\rm ext} = \gamma_{LV} \cos \theta$, requires a liquid-induced capillary pressure on the bottom of the wire (gray arrows) with an effective surface tension: (a) $\Gamma=-\gamma_{LV}\cos\theta$, (b) $\Gamma=0$, (c) $\Gamma=\gamma_{LV}$. Bottom: Displacement field $u(z)$ inside the wire due to immersion. Depending on the spatial distribution of capillary forces, these result into (a) a stretching (b) no deformation, or (c) a compression of the immersed part of the wire.}
\label{fig:Models}
\vspace{-6mm}
\end{figure}

In this Letter we demonstrate that when immersed in a liquid, curved surfaces of elastomeric solids experience a capillary pressure (or Laplace pressure). The magnitude of this \emph{solid capillary pressure} is determined by accurately measuring the deformation of a thin elastomeric wire suspended in a liquid reservoir (Fig.~\ref{fig:Models}). It is found that the immersed part of the wire is \emph{compressed}, consistent with a capillary pressure pushing on the wire from below -- this scenario is sketched in Fig.~\ref{fig:Models}c. The effective surface tension associated to this effect, $\Gamma$, is measured to be comparable to the liquid-vapor surface tension $\gamma_{LV}$, consistent with recent predictions~\cite{DMAS11}.  The key finding of our paper, however, is that the experiments reveal a highly unexpected force transmission at the contact line: the liquid-vapor surface tension is not pulling along the interface, but the force on the solid is oriented towards the interior of the liquid (Fig. 1c).

The existence of a capillary pressure at the solid-liquid interface has striking consequences. While the total force on a partially immersed elastic wire is easily measured as the external force $F_{\rm ext}$, it has remained unclear how this force is distributed along the wire~\cite{MWSA11,GBQ04,L61,R78,SG87,CGS96,LAL96,PBBB08,PRDBRB07,NRGB07,DMAS11}. Thermodynamics dictates that $F_{\rm ext}=\gamma_{LV} \cos \theta$ per unit contact line~\cite{GBQ04,MWSA11}, where $\theta$ is the contact angle of the liquid -- this principle is widely used to measure the liquid-vapor surface tension. 

However, the literature on the spatial transmission of this resultant thermodynamic force can be divided into three distinct scenarios. Figure~\ref{fig:Models}a: The contact line region exerts a purely normal force on the solid~\cite{L61,R78,SG87,CGS96,LAL96,PBBB08}, with no component parallel to the solid surface. Thermodynamic consistency with the vertical force $F_{\rm ext}$ requires a capillary pressure, pulling downward, localized in the curved region at the bottom of the wire ($\Gamma=-\gamma_{LV} \cos \theta)$.  Figure~\ref{fig:Models}b: The contact line region exerts a force parallel to the liquid-vapor interface~\cite{PBBB08,PRDBRB07,NRGB07}. The downward component parallel to the interface is exactly $\gamma_{LV}\cos \theta$, hence there is no capillary pressure at the bottom Laplace pressure ($\Gamma=0$). Figure~\ref{fig:Models}c: There is an upward capillary pressure at the bottom of the wire, with effective surface tension $\Gamma=\gamma_{LV}$~\cite{DMAS11}. Thermodynamic consistency is recovered only when the  force near the contact line has a downward parallel component $\gamma_{LV}(1+\cos \theta)$. 

The correct scenario for force transmission cannot be inferred from either macroscopic or mesoscopic (i.e. introducing disjoining pressure effects) calculations of the liquid free energy: before applying the virtual work principle to the solid, one needs to include the contribution of solid deformations to the free energy. This issue has been properly addressed in the case of crystalline solids \cite{MS04}, but not yet for soft solids. Experimentally, however, the elastic deformation of the wire provides a clear answer: the immersed part of the wire is either (a) stretched, (b) unaffected, or (c) compressed with respect to its dry reference state. The experiments described below show a clear compression for soft elastomers, quantitatively consistent with the third scenario. Therefore, we indeed find that the force on the solid near contact line is directed towards the interior of the liquid -- this is the central finding of this Letter. 
\begin{figure}[t!]
\includegraphics{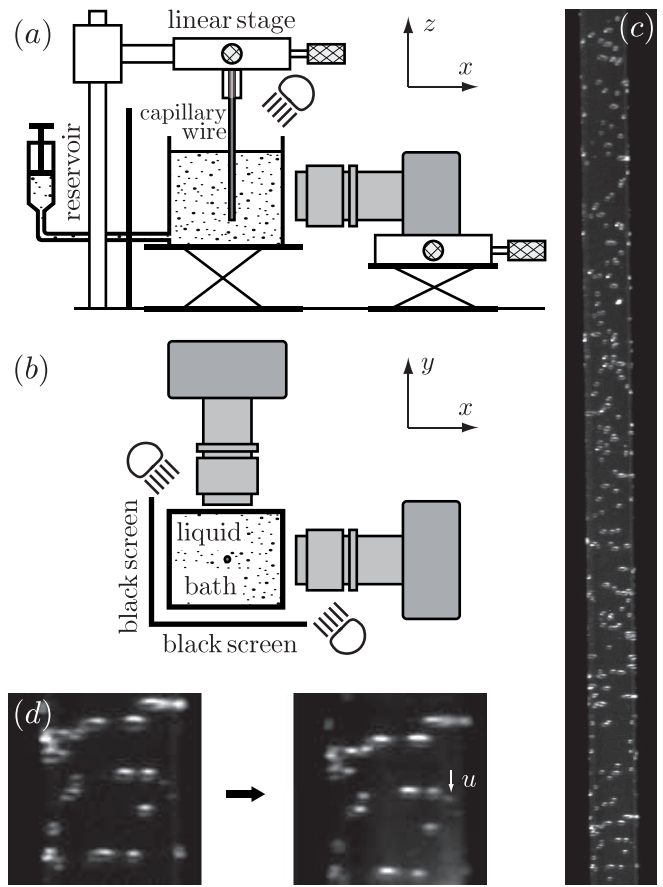}
\vspace{-6mm}
\caption{Schematic representation of the experiment: (a) side view and (b) top view. The vertical axis is denoted $z$, with $z=0$ at the liquid free surface. The deformation is characterized by a vertical displacement field $u(z)$, determined from markers in the wire (by convention, $u(0)=0$). The horizontal axis are noted $x$ and $y$, with $x=0$ and $y=0$ at the intersection of the camera optical axes. One camera is used to accurately calibrate the scale of the other. (c-d) Images of the wire with markers used to determine the displacement $u$.}
\vspace{-6mm}
\label{fig:Schematic}
\end{figure}

\textit{Experimental setup~--~}The experimental setup is depicted in Fig.~\ref{fig:Schematic}. The elastic wire is a filament made of low modulus polyvinylsiloxane elastomer (RTV EC00). Glass capillary tubes (Ringcaps $5~\mu$L) are filled with a mixture of polymer and curing agent containing dispersed polystyrene beads (Dynoseeds TS 20) of diameter $20~\mu$m, used as position markers. Once the polymer is cured, the glass capillary is cut and the filament is slid out of the capillary. With this procedure one finally obtains a cylindrical elastic rod of radius $R = 150~\mu$m and of length $20~$mm, glued at one end inside a $5~$mm piece of the glass capillary (Fig.~\ref{fig:Schematic}c). The immersion fluid is a $96\%$ ethanol. Its density $\rho_l=803\pm1\,{\rm kg\,m^{-3}}$ is measured at room temperature ($T=23.4^\circ $C) using a pycnometer of volume calibrated with ultrapure water ($18.1~{\rm M\Omega\,cm^{-1}}$). The liquid vapour surface tension is measured within $1\%$ with a tensiometer Kr\"uss MK100: $\gamma_{LV}=22.8~{\rm mN\,m^{-1}}$. The Young's modulus of the bulk elastomer is calibrated in the linear elastic regime (strain lower than $1\%$) to $E = 35$~kPa. In comparison to experiments on single drops \cite{JXWD11}, the present setup allows for a direct, robust test of the thermodynamic scenarios of Fig. \ref{fig:Models}.

The goal of the experiment is to measure the elastic deformation of the wire before and after the immersion. This is done by measuring the displacement of markers inside the wire (Fig.~\ref{fig:Schematic}d). The immersion is regulated by changing the level of the liquid, while the wire is held at the same location. The wire is imaged by two cameras Nikon D300 ($4288\times2848$ pixels,  $16$~bits raw images) mounted with extension tubes and macro-lenses, positioned at $90^\circ$ on an optical table. The absolute scale is then around $2~{\rm pix\,\mu m^{-1}}$. Focussing is controlled by translating each camera with a linear stage. The crucial step for accurate resolution of the displacements of the markers is the calibration of relative scales between an empty and filled container. This calibration is achieved by printing a $12 000$~dpi test pattern composed of an alternate array of $60~\mu$m wide black and white strips. From the correlation function between two images of the pattern we detemine the relative scale within $0.1 \%$. The local displacement of polystyrene particles is obtained within $1$ pixel by cross-correlation of images (Fig.~\ref{fig:Schematic}d).

\textit{Elasto-capillary derivation~--~}
The elastic deformation of the wire, characterized by the vertical displacement field $u(z)$, depends on the spatial distribution of the capillary stresses. The reference state for these displacements is the freely suspended wire submitted to its own weight, not yet in contact with the liquid reservoir. Hence, $u(z)$ probes only the liquid-induced stresses after immersion of the wire. Above the contact line, $z>0$, where the wire is still dry, the only stress is due to the external force balancing the thermodynamic force, $2\pi R F_{\rm ext}=2\pi R \gamma_{LV} \cos \theta$. This induces a vertical normal stress $\sigma_{zz} = 2\gamma_{LV}\cos \theta/R$, while the radial stress $\sigma_{rr}=0$. Below the contact line, $z < 0$, the radial stress exerted on the side walls of the wire consists of the hydrostatic pressure inside the liquid and a solid capillary pressure, 
\begin{equation}
\sigma_{rr}=\rho_l g z-\Gamma/R,
\label{eq:sig_rr_BC}
\end{equation}
where $\rho_l$ is the density of the liquid. In analogy to the Laplace pressure jump on a liquid-vapor interface, we hypothesize that the solid capillary pressure is proportional to the curvature of the solid-liquid interface. The unknown, effective surface tension $\Gamma$ is the central object of this Letter. Similarly, the normal stress on the bottom of the wire, i.e., at $z=-L$, reads
\begin{equation}\label{eq:sig_zz_BC}
\sigma_{zz} = - \rho_l g L - 2 \Gamma/R,
\end{equation}
This is most easily seen when the bottom of the wire is a hemi-spherical cap of radius $R$. The corresponding pressure jump is then $2\Gamma/R$, where the factor 2 arises from the two identical principle curvatures. We emphasize, however, that (\ref{eq:sig_zz_BC}) is valid for arbitrary shapes of the edge of the wire, as long as $R \ll L$~\cite{footnote}.
\begin{figure}[t!]
\includegraphics{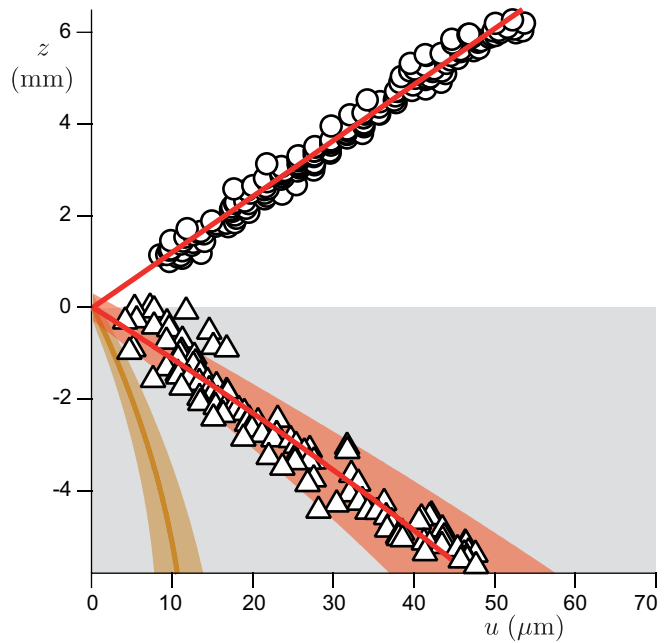}
\vspace{-6mm}
\caption{(color online) Local displacement $u(z)$ of the elastic wire compared to the situation when it is hanging from its own weight. Each point corresponds to the displacement of one polystyrene bead. Measurements above the contact line ($\circ$) and below ($\triangle$) are performed by rising the liquid level. Red lines are the best fit by equations (\ref{eq:airstrainrel}) and (\ref{eq:liquidstrainrel}). The calibration error bars are shown by the colored regions. The orange line shows the (compressive) contribution of hydrostatic pressure.}
\vspace{-6mm}
\label{fig:Compression}
\end{figure}

The displacement field induced by (\ref{eq:sig_rr_BC},\ref{eq:sig_zz_BC}) can be derived from standard elasticity. Like most elastomers, the material used in the experiment is almost incompressible: its Poisson ratio $\nu$ is such that $|\nu-1/2|\ll1$. Under this assumption, Hooke's law relates the stress tensor $\bar \sigma_{ij}$ to the strain tensor $\bar \varepsilon_{ij} $ and the pressure $P$ (which is the isotropic stress inside the solid) as~\cite{MZK02} $\bar \sigma_{ij} =  \frac{2}{3} E \bar \varepsilon_{ij} - P \delta_{ij}$ and  $\varepsilon_{ll}=0$.
%
%
Considering the limit where the radius $R$ is small compared to the length of the wire, the strain is homogeneous in a slice and depends only on the vertical coordinates $z$, i.e., $\epsilon_{zz}=u'(z)$, where $u(z)$ is the vertical displacement, and $\epsilon_{rr}=\epsilon_{\theta\theta}$. Radial displacements are much smaller than $u(z)$ by a factor $R/L\ll 1$. Using the incompressibility condition, one obtains $\epsilon_{zz}=u'(z)=-2\epsilon_{rr}=-2\epsilon_{\theta \theta}$. After eliminating $P$, one relates the vertical strain to the normal stress difference as $\epsilon_{zz}=(\sigma_{zz}-\sigma_{rr})/E$.
%

The resulting displacements along the wire are obtained by integration of the strain $\epsilon_{zz}$. Above the contact line, one finds a homogeneous stretching of the wire as sketched in Fig.~\ref{fig:Models}, 
\begin{equation}
\label{eq:airstrainrel}
u(z)=\frac{2 \gamma_{LV} \cos \theta \, z}{ER},\quad{\rm for}\quad z>0.
\end{equation}
Below the contact line, the stresses (\ref{eq:sig_rr_BC},\ref{eq:sig_zz_BC}) yield
\begin{equation}
u(z)=-\frac{\Gamma \, z}{ER} -\frac{\rho_l g z (z+2L)}{2E},\quad{\rm for}\quad z<0.
\label{eq:liquidstrainrel}
\end{equation}
Hence, there is a linear contribution due to the solid capillary pressure. Depending on the sign of $\Gamma$, this corresponds to compression ($\Gamma > 0$, Fig.~\ref{fig:Models}c) or to stretching ($\Gamma < 0$, Fig.~\ref{fig:Models}a) in the vertical direction. The strain $u'(z)$ thus provides a direct  measurement of the sign and magnitude of the effective surface tension $\Gamma$. 

\textit{Results and Discussion~--~} Figure~\ref{fig:Compression} shows the displacement field obtained after immersion of a homogeneous elastic wire. As expected, the displacements in the air ($z>0$, circles) correspond to a homogeneous stretching of the wire. By contrast, we systematically observed a compression of the submerged part of the wire ($z<0$, triangles). This is consistent with the scenario proposed in Fig.~\ref{fig:Models}c, with a positive, nonzero surface tension $\Gamma>0$. Quantitatively, our data indeed agrees with the recent prediction $\Gamma=\gamma_{LV}$~\cite{DMAS11}. Above the contact line, the best fit of the data by equation (\ref{eq:airstrainrel}) determines the dimensionless parameter $\frac{\gamma_{LV} \cos \theta}{ER}=4.03\times10^{-3}$, within 2\%. As there is no optical scaling factor in this case, possible errors only result from the detection of markers. In fact, measuring the contact angle using a photograph ($\cos\theta=0.7\pm0.2$) limits the accuracy. The dimensionless elastocapillary parameter $\frac{\gamma_{LV}}{ER}$ is thus around $6\pm2\times10^{-3}$, consistent with a calibrated value from a separate determination of the Young's modulus ($4.4\times10^{-3}$). Below the contact line, the best fit of the data with equation (\ref{eq:liquidstrainrel}), including the hydrostatic contribution, gives $\frac{\Gamma}{ER}=6.4\times10^{-3}$, within 15\%. So indeed,
\begin{equation}
\frac{\Gamma}{\gamma_{LV}}=1.2\pm0.3.
\end{equation}
\begin{figure}[t!]
\includegraphics{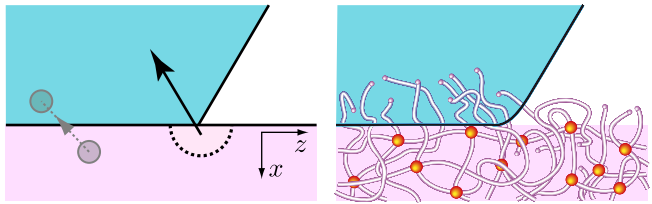}
\vspace{-6mm}
\caption{(color online) (a) Schematic showing the origin of the tangential force component exerted by the entire liquid (blue wedge) on the solid near to the contact line, due to attraction of solid by liquid elements (grey arrow). (b) Sketch of the polymers at the free surface of the elastomer, as they are pulled towards the interior of the liquid.}
\vspace{-6mm}
\label{fig:Elastomer}
\end{figure}

These findings reveal that the capillary force near the contact line is neither normal to the solid, nor parallel to the liquid-vapour interface, but is oriented toward the wedge of liquid: as sketched in Fig.~\ref{fig:Models}c, thermodynamics is only recovered when the force on the solid has a (downward) tangential component $F_t = \Gamma+\gamma_{LV}\cos\theta \approx \gamma_{LV}(1+\cos\theta)$. This remarkable result can be explained from the long-range van der Waals attractions of the liquid molecules, pulling on the solid molecules near the surface. The geometry near the contact line induces a resultant force that is biased towards the direction of the liquid domain, where most of the liquid molecules are located (Fig.~\ref{fig:Elastomer}a). This bias can be computed exactly when assuming that the liquid and solid constitute homogeneous phases that are mutually attractive. Expressing the liquid potential energy as $\pi_{ls}$, one obtains the force density $-\nabla \pi_{ls}$ inside the solid. The tangential force on the solid follows from integration over the solid domain, $F_t = \int dx dz \,\left[ - \partial_z \pi_{ls}\right] = \int dx \, \pi_{ls}^*$, where $\pi_{ls}^*$ is the liquid potential far away from the contact line. A further simplification arises since away from the contact line the liquid domain is a semi-infinite phase: $\pi_{sl}^*$  is only a function of $x$, the distance to the solid-liquid interface. It is known that the $x$-integral over $\pi_{ls}^*$ is normalized in terms of the surface tensions as $\gamma_{SL}-\gamma_{SV}-\gamma_{LV} =-\gamma_{LV}(1+\cos\theta)$~\cite{KM1991,DMAS11,GD1998}, regardless of the type of microscopic interaction. Hence, our mean field model based on homogeneous attracting phases quantitatively captures the observed $F_t$, and supports our experimental result that the contact line force on the solid is directed towards the interior of the liquid.

From a broader perspective, our work demonstrates that the details of elasto-capillary interactions cannot be captured by macroscopic thermodynamic arguments and requires microscopic modeling. 
The surface of an elastomer consists of free flexible polymers which are attracted by the liquid, as shown in Fig.~\ref{fig:Elastomer}b. The left-right symmetry of the free chains is broken in the vicinity of the contact line, resulting in a pulling force that is transmitted towards the bulk of the elastomer. The transmission of such a tangential capillary force is specific to a solid interface: a liquid would be unable to sustain such a shear as it is able to rearrange its molecules.

\begin{acknowledgments}
We thank D. Bartolo, J. Bico, P. M\"uller, E. Rapha\"el, B. Roman and J. Sprakel for valuable discussions.
\end{acknowledgments}


\begin{thebibliography}{}
\bibitem{MS04} P. Muller, and A. Sa\'ul Surf. Sci. Rep. {\bf 54}, 157 (2004).
\bibitem{BGIE97} C.E. Bach, M. Giesen, H. Ibach, and T.L. Einstein, Phys. Rev. Lett. {\bf 78}, 4225 (1997).
\bibitem{FF97} A. Filipetti and V. Fiorentini, Surf. Sci. {\bf 377}, 112 (1997).
\bibitem{WK77} P. Wynblatt and R. Ku, Surf. Sci. {\bf 65}, 511 (1977).
\bibitem{I04} H. Ibach, Surf. Sci. {\bf 556}, 71 (2004). 
\bibitem{AVMJ88} O.L. Alerhand, D. Vanderbilt, R.D. Meade and J.D. Joannopoulos, Phys. Rev. Lett. {\bf 61}, 1973 (1988). 
\bibitem{MSM05} J. J. M\'etois, A. Sa\'ul, and P. Muller, Nat. Mater. {\bf 4}, 238 (2005). 
\bibitem{R02} S. Rousset, Mater. Sci. Eng. B {\bf 96}, 169 (2002). 
\bibitem{MZK02} J.-P.~Mercier, G.~Zambelli, W.~Kurz, \textit{Introduction to materials science} (Elsevier, Paris, 2002)
  \bibitem{SG87} M. E. R. Shanahan and P. G. de Gennes, Adhesion 11 (Elsevier Applied Science, London, 1987).
  \bibitem{CGS96} A. Carre, J.-C. Gastel, and M. E. R. Shanahan, Nature {\bf 379}, 432 (1996).
\bibitem{MPFPP10}  S.~Mora,  T.~Phou,  J.-M.  Fromental,  L.~M. Pismen, and  Y.~Pomeau,  Phys. Rev. Lett. {\bf 105}, 214301 (2010). 
\bibitem{JXWD11}  E.~R. Jerison,  Y.~Xu,   L.~A. Wilen, and  E.~R. Dufresne, Phys. Rev. Lett. {\bf106}, 186103  (2011). 
  \bibitem{DMAS11} S.~Das, A.~Marchand, B.~Andreotti, and J.~H. Snoeijer Phys. Fluids. {\bf 23}, 072006 (2011).
  \bibitem{GBQ04} P.-G. de Gennes, F. Brochard-Wyart, and D. Quere, \textit{Capillarity and wetting phenomena: drops, bubbles, pearls, waves} (Springer, New York, 2004).
   \bibitem{L61} G. R. Lester, J. Colloid Sci. {\bf 16}, 315 (1961).
\bibitem{R78} A. I. Rusanov, J. Colloid Inter. Sci. {\bf 63} 330 (1978).
  \bibitem{LAL96} D. Long, A. Ajdari, and L. Leibler, Langmuir {\bf 12}, 5221 (1996).
  \bibitem{PBBB08}  R. Pericet-Camara, A. Best, H. J. Butt and E. Bonaccurso, Langmuir {\bf 24}, 10565 (2008).
  \bibitem{PRDBRB07} C.~Py, P.~Reverdy, L.~Doppler, J.~Bico,  B.~Roman, and C.~N. Baroud, Phys. Rev. Lett. \textbf{98}, 156103 (2007). 
\bibitem{NRGB07} S.~Neukirch, B.~Roman,  B.~de~Gaudemaris, and J.~Bico, J. Mech. Phys. Sol. \textbf{55}, 1212 (2007). 
\bibitem{MWSA11} A. Marchand, J. Weijs, J. H. Snoeijer, and B. Andreotti, Am. J. Phys. {\bf 79}, 999 (2011).
\bibitem{footnote} The total force exerted on the bottom is obtained by integrating over the curved surface, $d\mathbf{F} = \Gamma \kappa d\mathbf{S}$. As the curvature $\kappa$ is a derivative of the tangent vectors, the integrated force is independent on the surface shape~\cite{MWSA11}.   
\bibitem{KM1991} J. B. Keller and G. J. Merchant, J. Stat. Phys. {\bf 63}, 1039 (1991).
\bibitem{GD1998} T. Getta and S. Dietrich, Phys. Rev. E {\bf 57}, 655 (1998).
\end{thebibliography}
\end{document}